\documentclass[sigconf, nonacm]{acmart}

\usepackage{amsmath,amsfonts}
\usepackage{graphicx}
\usepackage{textcomp}
\usepackage{xcolor}
\usepackage{url}
\usepackage{booktabs}

\newcommand{\project}{\pi}
\newcommand{\select}{\sigma}
\newcommand{\reljoin}{\bowtie}
\newcommand{\group}{\gamma}
\newcommand{\bagcup}{\uplus}

\begin{document}

\title[Co-Usage Patterns of Programming Languages on Stack Overflow]{Programming Language Co-Usage Patterns on Stack Overflow:\\
Analysis of the Developer Ecosystem}

\author{Bachan Ghimire}
\email{bachan48@uvic.ca}
\affiliation{%
  \institution{University of Victoria}
  \city{Victoria}
  \state{BC}
  \country{Canada}
}

\author{Nitin Gupta}
\email{ngupta@uvic.ca}
\affiliation{%
  \institution{University of Victoria}
  \city{Victoria}
  \state{BC}
  \country{Canada}
}

\renewcommand{\shortauthors}{Ghimire and Gupta}

\begin{abstract}
Understanding how developers combine programming languages in practice
reveals the hidden structure of the software ecosystem: which
languages are used as complements, which define coherent technology
stacks, and which bridge disparate communities. We present a
three-phase empirical pipeline that mines Stack Overflow posts by
hundreds of thousands of developers across 186 programming languages,
applying FP-Growth frequent itemset mining, Latent Dirichlet
Allocation topic modeling, and Louvain community detection on a
weighted co-usage graph, with the goal of characterizing co-usage
coupling, latent developer specializations, and macro-level ecosystem
structure simultaneously from behavioral data. FP-Growth identifies
tight coupling clusters such as shell/bash, Swift/Objective-C, and
the C-family with lift values far exceeding what individual language
popularity predicts. LDA produces 25 developer profiles including
Apple-platform developers, scientific and hardware programmers,
functional/academic programmers, and two distinct Unix scripting
sub-profiles. Louvain partitions the language graph into three
macro-communities: web/enterprise, Apple ecosystem, and
systems/scientific, and identifies Java as the highest-degree hub
connecting all three. All three methods independently converge on
the same ecosystem structure, providing strong cross-method
validation of the findings. 
\end{abstract}




\maketitle

% ------------------------------------------------------------------
\section{Introduction}

The number of actively used programming languages has grown
substantially over the past two decades. Modern software systems are
rarely built in a single language: a data science pipeline might
combine Python for modeling, SQL for data access, and Bash for
orchestration; an iOS application typically requires both Swift for
new code and Objective-C for legacy components; a high-performance
scientific application may span C, Fortran, and Python. Understanding
which languages are systematically used in combination, and which
combinations are incidental versus structurally coupled, has broad
practical importance for tool builders, language designers,
documentation authors, and educators.

Stack Overflow\footnote{\url{https://stackoverflow.com}} is the
largest publicly available record of developer behavior at scale.
Its question-and-answer corpus, tagged by users with programming
language names, provides a fine-grained proxy for individual language
activity: a developer who asks or answers questions about Python and
SQL is, with high probability, one who actively uses both. Aggregated
across hundreds of thousands of users and 36 million posts, this
record encodes the collective structure of the developer ecosystem
in a form amenable to large-scale empirical analysis.

Prior work on Stack Overflow has characterized individual language
communities~\cite{chakraborty2021}, tracked temporal evolution of
technology adoption~\cite{moutidis2021,mondal2023}, studied the
difficulty of learning a new language after knowing
another~\cite{shrestha2020}, and analyzed security challenges across
different languages~\cite{croft2021}. Separately, studies on
multi-language development have examined interoperability
tools~\cite{cherny2024} and the structural properties of language
dependency ecosystems~\cite{blanthorn2019}. However, no study has
applied a multi-method pipeline to characterize \emph{co-usage
coupling} across the entire active language landscape at the
individual-user level, distinguishing tight structural dependencies
from coincidental co-occurrence driven by popularity.

This paper fills that gap with three complementary analyses. FP-Growth
frequent itemset mining identifies language pairs and triples whose
co-usage probability exceeds what their individual popularity
explains. LDA topic modeling surfaces latent developer profiles by
treating each developer's language basket as a document. Louvain
community detection on a weighted co-usage graph provides a
macro-level structural view, partitioning the language space into
communities and identifying hub languages that bridge specializations. Three research questions guide the analysis:

\smallskip
\noindent\textbf{RQ1.} \textit{Which programming language combinations
exhibit co-usage coupling beyond what individual language popularity
explains, and what do these patterns reveal about structural language
dependencies?}

\smallskip
\noindent\textbf{RQ2.} \textit{What latent developer profiles emerge
from multi-language usage patterns, and do these correspond to
recognizable technology domains?}

\smallskip
\noindent\textbf{RQ3.} \textit{How is the co-usage graph structured at
the community level, and which languages function as cross-community
hubs?}

The main contributions are: (1)~association rules quantifying structural language
coupling, separating it from popularity effects via lift; (2)~25
empirically derived developer profiles partitioning the Stack Overflow
population into interpretable specializations; (3)~a weighted co-usage
graph and community structure providing a macro-level ecosystem map;
and (4)~a convergence result showing that three methodologically
independent approaches identify the same ecosystem boundaries.

% ------------------------------------------------------------------
\section{Related Work}

\subsection{Mining Stack Overflow for Developer Behavior}

Stack Overflow has been extensively mined to understand developer
activity. Ahmad et al.~\cite{ahmad2018} surveyed SO mining across
hundreds of studies. The platform has been used to study developer
challenges in Docker~\cite{haque2020}, refactoring~\cite{peruma2021},
low-code development~\cite{alamin2021}, and continuous
integration~\cite{zahedi2020}, typically through LDA on question and
answer text.

Studies targeting programming languages on SO have examined how
developers discuss new languages such as Go, Swift, and
Rust~\cite{chakraborty2021}, finding a slow initial growth phase
before community support accelerates. Croft et al.~\cite{croft2021}
analyzed security challenges across 15 languages using topic
modeling, finding substantial variation by language. Cagnoni et
al.~\cite{cagnoni2020} found that developer sentiment varies
systematically by language. Shrestha et al.~\cite{shrestha2020}
studied why learning a second language is difficult after knowing
one, finding widespread interference from prior language knowledge.
These studies characterize individual language communities but do
not analyze co-usage relationships between languages.

\subsection{Community Structure and Ecosystem Analysis}

Moutidis and Williams~\cite{moutidis2021} analyzed community
evolution on Stack Overflow from 2008 to 2020, finding that web
development communities are the largest and most persistent, with
little cross-community user movement. Their study operates at the
year-by-year level and treats languages and frameworks together;
ours focuses exclusively on programming languages at the
individual-user co-usage level. Antelmi et al.~\cite{antelmi2023}
examined developer community formation around the 20 most popular
languages on Stack Overflow and Reddit using hypergraph analysis.
Neither study characterizes structural coupling between languages
via association rule mining or models individual developer profiles.

\subsection{Multi-Language Development}

Cherny-Shahar and Feitelson~\cite{cherny2024} conducted the most
directly related prior study, analyzing multi-language
interoperability across 414,486 GitHub repositories and 22 million
Stack Overflow questions, with focus on interoperability tools and
C as a central pillar. Our work differs in unit of analysis, we
focus on per-user co-usage patterns rather than per-project language
composition, and in methodology, applying three complementary
analytical lenses rather than cataloging interoperability tools.

Mondal et al.~\cite{mondal2023} analyzed technology usage span on
Stack Overflow, finding C\# and Java have the highest average span
per developer. Blanthorn et al.~\cite{blanthorn2019} used tensor
decomposition to compare library-level dependency ecosystems across
five language communities.

\subsection{Developer Profiles and Specialization}

Blanco et al.~\cite{blanco2020} studied the Java community on Stack
Overflow over 10 years using graph mining. Mukta et
al.~\cite{mukta2024} predicted language preferences from personality
traits. Survey-based approaches~\cite{durdev2024} capture only
self-reported behavior.

\subsection{Research Gap}

No prior study has applied a multi-method pipeline to characterize
co-usage coupling, latent developer profiles, and ecosystem community
structure simultaneously from per-user activity traces on Stack
Overflow. The co-usage lens which languages individual developers
use in combination, and how strongly is distinct from language
popularity and from project-level multi-language composition.

% ------------------------------------------------------------------
\section{Methodology}

\subsection{Dataset}

We use a publicly available Stack Overflow data
dump comprising posts
through 2024. Four tables are used: \texttt{posts} (36.1M rows),
\texttt{poststags} (41.6M rows), \texttt{users} (7.25M rows), and
a curated \texttt{languages} list of 262 valid programming language
tag names used to filter out framework, OS, and tool tags.
Table~\ref{tab:datasets} summarizes the raw data.

\begin{table}[h]
  \caption{Raw dataset files}
  \label{tab:datasets}
  \begin{tabular}{lrr}
    \toprule
    File & Rows & Key columns \\
    \midrule
    posts      & 36,149,134 & postid, posttypeid, parentid, userid \\
    poststags  & 41,605,024 & postid, tag \\
    users      &  7,250,739 & userid \\
    languages  &        262 & language \\
    \bottomrule
  \end{tabular}
\end{table}

\subsection{Processing Pipeline}

We implement a SQL-based pipeline that transforms raw posts into
per-user language baskets, the set of languages a user has engaged
with across their Stack Overflow history. Both questions
(\texttt{posttypeid}=1) and answers (\texttt{posttypeid}=2) carry
language signal. Answers reference their parent question via
\texttt{parentid} to inherit the question's tags. We join the
unified post table against the language list, retaining only
recognized programming language tags. We then count per-user
language occurrences and apply two filters: (i)~drop any
(user, language) pair with usage count of one, removing incidental
encounters; and (ii)~drop any user with fewer than two qualifying
languages, since single-language users carry no co-usage signal.
The pipeline is formalized in relational algebra as follows.

\medskip
\noindent\textbf{\textit{post\_user\_languages}:}
\begin{align*}
&\project_{\mathit{pid},\mathit{uid},\mathit{tag}}\bigl(
  \select_{\varphi_Q}(
    \mathit{ptags}
    \reljoin_{\mathit{ptags.pid}=q.\mathit{pid}}
    \rho_{q}(\mathit{posts})
  )
\bigr) \\
&\bagcup_{\!\text{all}} \\
&\project_{\mathit{pid},\mathit{uid},\mathit{tag}}\bigl(
  \select_{\varphi_A}(
    \mathit{ptags}
    \reljoin_{\mathit{ptags.pid}=a.\mathit{parid}}
    \rho_{a}(\mathit{posts})
  )
\bigr)
\end{align*}

\noindent where $\varphi_Q$:
$\mathit{tag}\!\in\!\project_{\mathit{lang}}(\mathit{langs})
\wedge \mathit{uid}\!\sim\!\texttt{\^{}{\char`\\}d+\$}
\wedge \mathit{pid}\!\sim\!\texttt{\^{}{\char`\\}d+\$}$;
and $\varphi_A$: $\varphi_Q
\wedge \mathit{parid}\!\sim\!\texttt{\^{}{\char`\\}d+\$}$

\medskip
\noindent\textbf{\textit{user\_results}:}
\[
\select_{\mathit{cnt}>1}\!\left(
  \group_{\mathit{uid},\mathit{tag};\,\mathit{cnt}(*)\to c}(\mathit{ul})
\right)
\]

\medskip
\noindent\textbf{\textit{final\_query:}}
\[
\select_{\mathit{cnt}(\mathit{uid})>1}\!\left(
  \group_{\mathit{uid};\,\mathit{agg}(\mathit{tag})\to\mathit{langs}}(\mathit{ur})
\right)
\]

Table~\ref{tab:pipeline} shows the effect of the filters applied.
The 82\% reduction from 2.38M to 435k users show the
majority of Stack Overflow users touched language tags only
incidentally. The 435k retained users have demonstrated, recurrent
multi-language activity.

\begin{table}[h]
  \caption{Pipeline output statistics}
  \label{tab:pipeline}
  \begin{tabular}{lr}
    \toprule
    Metric & Value \\
    \midrule
    Total users with any language activity & 2,378,706 \\
    Users after filtering                  &   435,803 \\
    Unique languages in baskets            &       186 \\
    \bottomrule
  \end{tabular}
\end{table}

\subsection{Analysis Methods}

\subsubsection{FP-Growth -- Co-Usage Coupling}
Frequent itemset mining~\cite{agrawal1994} identifies all item sets
whose \emph{support}, the fraction of baskets containing the set
-- exceeds a threshold $\sigma$. Association rules $A \Rightarrow B$
additionally require minimum \emph{confidence} $P(B \mid A) \geq
\gamma$. The classic Apriori algorithm enumerates candidates
level-by-level, generating an exponential number of candidate sets
on dense data. FP-Growth~\cite{han2000} avoids this by compressing
the transaction database into a compact prefix tree (FP-tree) and
mining patterns directly from that structure without candidate
generation, making it tractable at the scale of our 435,803-basket
dataset. \emph{Lift} is the primary metric for RQ1:
\[
  \mathrm{lift}(A \Rightarrow B)
  = \frac{P(A \cup B)}{P(A) \cdot P(B)}.
\]
Lift measures how much more likely $B$ is given $A$ relative to
$B$'s unconditional probability. Lift of 1 means the two languages
co-occur at exactly the rate their individual prevalences predict
-- statistical independence. Lift substantially above 1 indicates
genuine structural coupling far more frequent than either language's
popularity alone can explain. A naive co-occurrence count cannot
distinguish two languages that appear together simply because both
are popular (Python and Java each appear in 20--25\% of baskets, so
their raw co-occurrence count is large even under independence) from
two that appear together because of structural necessity (Swift and
Objective-C, which co-occur at 8.35 times the independent rate).
Lift directly controls for this confound. An alternative, Pearson
or Spearman correlations between binary language vectors, is
mathematically equivalent to the Phi coefficient and suffers the
same popularity inflation. We use the \texttt{mlxtend}
library~\cite{raschka2018} with $\sigma = 0.02$ and $\gamma = 0.5$.

\subsubsection{LDA -- Developer Profile Discovery}
LDA~\cite{blei2003} models each document as a mixture of $k$ latent
topics, each a probability distribution over words. We treat each
user's language basket as a bag-of-languages document: languages are
\emph{words} and users are \emph{documents}. LDA simultaneously
infers the per-user topic mixture and the per-topic language
distribution. The choice of LDA over \emph{k-means clustering} deserves explicit
justification. K-means minimizes sum-of-squared Euclidean distances
between data points and cluster centroids. Applied to language
baskets, this requires embedding each user as a binary vector in
$\mathbb{R}^{186}$ and treating that vector as a geometric point.
This is fundamentally inappropriate for two reasons.

First, Euclidean distance imposes magnitude and ordering that do not
exist between categorical items. No language is \emph{greater than}
or \emph{between} any other. The Euclidean midpoint between a
Python-user and a Java-user is a user who is \emph{half a
Python-user and half a Java-user}, a concept that provides no meaning in the
categorical domain. The metric treats the presence of an obscure
language such as Prolog identically to Python in its contribution
to inter-user distance, despite the two carrying vastly different
profile signals. Second, k-means forces hard cluster assignment: every user belongs
to exactly one cluster. But developers routinely span multiple
specializations. A researcher who writes Python scripts, Bash
automation, and C++ performance kernels genuinely inhabits multiple
profiles simultaneously. LDA's \emph{soft membership} models this
directly: each user receives a probability distribution over
profiles, and their basket is drawn from a mixture of topics
proportional to those probabilities. As Silva et al.~\cite{silva2021}
document in their survey of topic modeling in software engineering
research, LDA is the dominant choice in SE empirical studies
precisely because of this soft membership property and its
production of human-interpretable topics from developer activity.

We use scikit-learn's batch variational Bayes with $k = 25$,
random seed 0, and 20 E-M iterations. The choice of $k = 25$ was
guided by manual inspection of interpretability across $k \in
\{15, 20, 25, 30\}$; at $k = 25$ topics are non-redundant and
each loads clearly on 2--5 languages. Topic labels are assigned
manually following the convention of Silva et al.~\cite{silva2021}.

\subsubsection{Louvain Community Detection -- Ecosystem Structure}
We construct a weighted, undirected co-usage graph
$G = (V, E, \omega)$ where $V$ is the set of 186 languages, an
edge $(u,v) \in E$ exists for every co-occurring language pair, and
$\omega(u,v)$ counts users who have both in their basket. The graph
has 6,502 edges. We load it into
Neo4j\footnote{\url{https://neo4j.com/}} for storage and querying,
and apply the Louvain algorithm~\cite{blondel2008} via
\texttt{python-louvain} on the weighted adjacency. The Louvain algorithm optimizes modularity $Q$~\cite{newman2004}:
\[
  Q = \frac{1}{2m} \sum_{i,j}
  \left[ A_{ij} - \frac{k_i k_j}{2m} \right]
  \delta(c_i, c_j),
\]
where $A_{ij} = \omega(i,j)$, $k_i = \sum_j A_{ij}$ is the weighted
degree of node $i$, $m = \frac{1}{2}\sum_{ij} A_{ij}$ is the total
edge weight, and $\delta(c_i, c_j)$ is 1 if nodes $i$ and $j$ share
a community. Modularity measures edge density within communities
relative to a degree-preserving random graph. Louvain optimizes $Q$
through a two-phase greedy process: each node is moved to the
neighboring community yielding the greatest local $Q$ gain; then
communities collapse into super-nodes and the process repeats until
convergence. The choice of Louvain over \emph{spectral clustering} or
\emph{DBSCAN} is deliberate. Spectral clustering would project the
graph Laplacian into a $d$-dimensional eigenspace and then run
k-means, reintroducing the hard-assignment and categorical-distance
problems discussed above, while additionally discarding topological
signals in the projection. DBSCAN requires a meaningful distance
metric between language nodes; as argued, no such metric exists for
categorical items. Louvain operates directly on the weighted
adjacency structure, preserving the full pairwise co-usage signal
without projection or dimensionality reduction, and scales
efficiently to our 186-node graph.

% \subsection{Methodological Justification}

% The three methods address different structural properties and are
% not interchangeable. FP-Growth identifies \emph{local, directional,
% discrete associations} with explicit statistical thresholds,
% producing falsifiable rules with quantified support, confidence, and
% lift -- but cannot model the continuous spectrum of partial
% membership real developers exhibit. LDA captures \emph{latent,
% soft, probabilistic profiles}, modeling developers as mixtures of
% specializations, accurately reflecting multi-disciplinary expertise
% -- but requires a fixed $k$ and produces undirected topics without
% confidence-ranked rules. Louvain captures \emph{global topological
% structure} from thousands of pairwise relationships simultaneously
% -- but produces hard community assignments and is sensitive to graph
% density.

% The scientific value of applying all three is cross-method
% convergence: a finding consistent across all three analyses -- each
% with different theoretical assumptions and different failure modes
% -- is substantially more credible than a finding from any single
% method alone.

% ------------------------------------------------------------------
\section{Results}

\subsection{RQ1: Language Co-Usage Coupling}

FP-Growth yields 106 frequent itemsets and 62 association rules.
Table~\ref{tab:fprules} shows the top rules by lift.

\begin{table}[h]
  \caption{Top association rules by lift}
  \label{tab:fprules}
  \begin{tabular}{llrrr}
    \toprule
    Antecedent & Consequent & Supp. & Conf. & Lift \\
    \midrule
    shell          & bash        & 0.025 & 0.610 & 11.06 \\
    swift          & objective-c & 0.033 & 0.670 &  8.35 \\
    c, c\#         & c++         & 0.027 & 0.746 &  4.20 \\
    c++, python    & c           & 0.031 & 0.514 &  4.04 \\
    c, python      & c++         & 0.031 & 0.661 &  3.72 \\
    c              & c++         & 0.078 & 0.613 &  3.45 \\
    \bottomrule
  \end{tabular}
\end{table}

\textbf{Shell/Bash coupling (lift 11.06).}
The strongest association is shell $\Rightarrow$ bash: 61\% of users
who engage with shell also engage with bash, at a lift of 11.06.
This is the highest coupling found because shell and bash are nearly
interchangeable in practice, bash being the predominant POSIX
shell implementation. The high lift confirms this is not a
popularity artifact: bash individually appears in 5.5\% of baskets,
so independence would predict only 5.5\% confidence; the 11x excess
reflects genuine functional coupling. This is consistent with
Mondal et al.'s~\cite{mondal2023} finding that bash has among the
highest technology usage span on Stack Overflow.

\textbf{Swift/Objective-C coupling (lift 8.35).}
The second-strongest coupling reflects the iOS/macOS development
context, where both languages coexist in production codebases:
Swift for new development, Objective-C for legacy components. The
8.35x lift confirms structural platform coupling, not voluntary
pairing. Chakraborty et al.~\cite{chakraborty2021} found Swift
among the fastest-growing languages on Stack Overflow; our result
adds that Swift adoption is almost always accompanied by
Objective-C co-usage.

\textbf{C-family coupling (lifts 3.45--4.20).}
The c $\Rightarrow$ c++ rule and multi-antecedent variants
(c + python $\Rightarrow$ c++; c + c\# $\Rightarrow$ c++) confirm
a persistent systems-programming pairing. The higher lift for
c + python $\Rightarrow$ c++ reflects the common pattern of Python
wrapping performance-critical C/C++ code. The c + c\# rule captures
developers spanning .NET and native code, common in Windows-platform
systems programming.

\textbf{Java as a cross-profile hub.}
Java appears as a consequent in nearly a dozen rules whose
antecedents span systems (c, c++), web (php, javascript), and
enterprise (c\#) languages with modest lifts (1.5--2.2). The
breadth of antecedents predicting Java's presence is unique in the
rule set, identifying Java as a common co-occurrence partner across
fundamentally different specializations. This hub property is
corroborated in Section~\ref{subsec:rq3}.

\textbf{Python and JavaScript breadth without coupling.}
Both languages appear frequently but at low lift (1.0--1.8),
reflecting near-independence: their co-occurrence with other
languages is approximately proportional to their individual base
rates. Python appears in 24\% of baskets and JavaScript in 29\%,
making them the two most prevalent languages in the dataset  They are
generic-utility languages, in contrast to the domain-specific
coupling of shell/bash, Swift/Objective-C, and the C family.

\begin{figure}[h]
  \centering
  \includegraphics[width=\columnwidth]{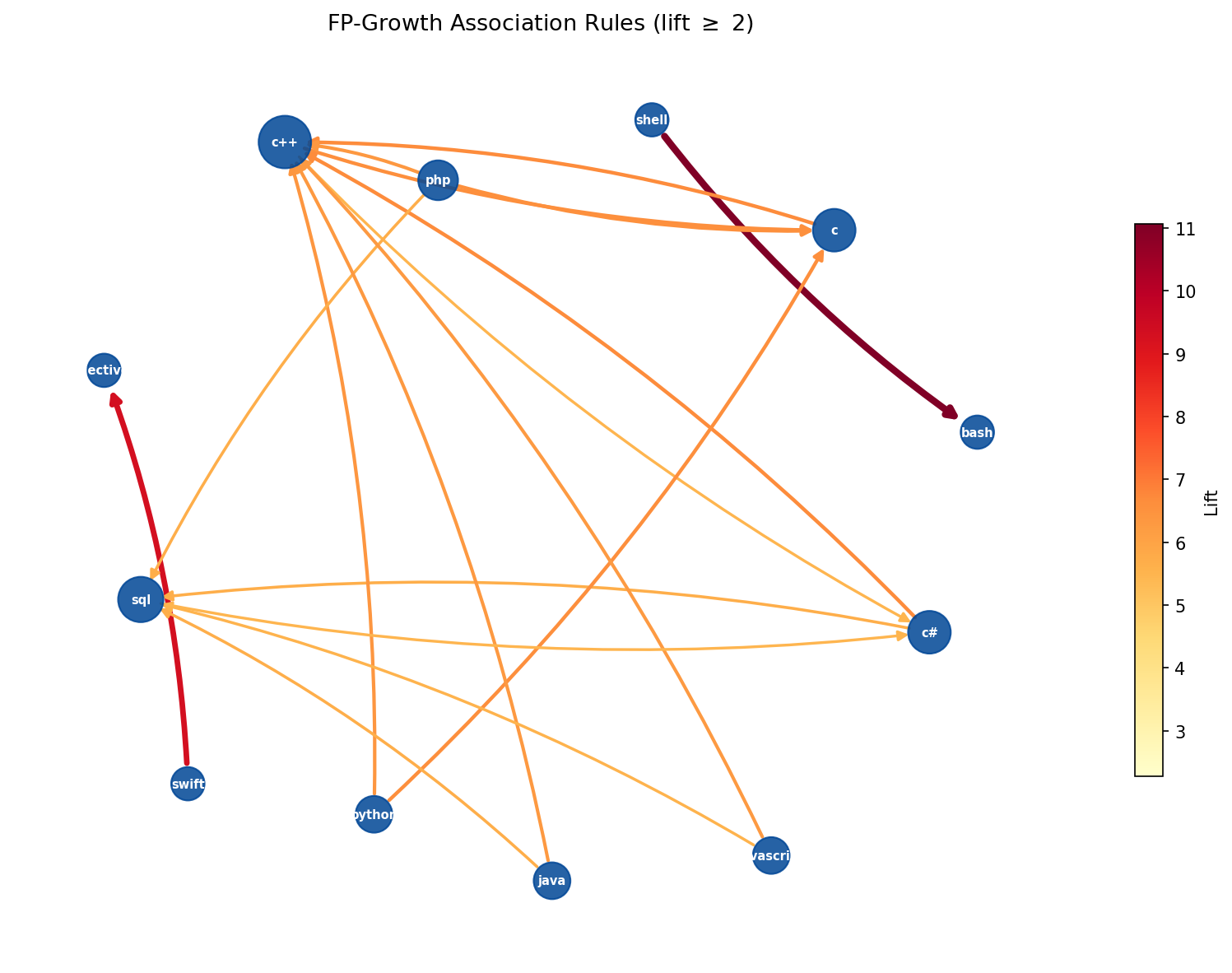}
  \caption{Association rule network (lift $\geq 2$). Edge color:
  yellow = low lift, red = high lift. Peripheral dyads are the
  tightest coupling pairs. The central cluster shows C-family rules
  converging on c++ and java.}
  \label{fig:fpnet}
\end{figure}

Fig.~\ref{fig:fpnet} visualizes rules with lift $\geq 2$. The
shell/bash and swift/objective-c dyads appear as isolated peripheral
clusters, visually confirming their self-contained nature. The
central component shows C-family rules converging on c++ and java.

\subsection{RQ2: Latent Developer Profiles}

LDA with $k = 25$ produces interpretable profiles, most loading
heavily on 2--4 languages with near-zero weight elsewhere.
Table~\ref{tab:lda} shows eight representative topics; the
remaining 17 cover Android/mobile, data science, and web
sub-stacks.

\begin{table}[h]
  \caption{Selected LDA developer profiles (top 5 languages each)}
  \label{tab:lda}
  \begin{tabular}{clr}
    \toprule
    Topic & Top languages & Users \\
    \midrule
     3 & perl, bash, sed, awk, python        & 10,781 \\
     9 & swift, objective-c                  & 15,933 \\
    11 & php, javascript, java               & 66,094 \\
    14 & haskell, lisp, scheme, rust, ocaml  &  5,644 \\
    17 & bash, shell, python                 & 20,007 \\
    20 & typescript, javascript              &  6,762 \\
    22 & python, fortran, vhdl, verilog      & 28,892 \\
    23 & powershell, vbscript                &  6,893 \\
    \bottomrule
  \end{tabular}
\end{table}

\textbf{Topic 9 -- Apple Platform Developer ($\sim$16k users).}
Nearly all weight concentrates on swift and objective-c, mirroring
the FP-Growth result independently. Corroboration across a
rule-based and a probabilistic method strengthens the finding. For
tool builders targeting iOS/macOS developers, this profile
establishes that the audience uses both languages simultaneously.

\textbf{Topics 3 and 17 -- Two Unix Scripting Sub-Profiles.}
A non-obvious finding: what appears as a single Unix scripting
community comprises two distinct LDA topics. Topic~3 is Perl-heavy
and tool-oriented (sed, awk), representing an older generation of
Unix scripters. Topic~17 is Python-heavy, representing the newer
generation that adopted Python as the primary scripting language.
User counts (10,781 vs 20,007) suggest the Python-heavy sub-profile
is twice as prevalent, consistent with Python's dominance in the
2010s. Shrestha et al.~\cite{shrestha2020} showed that Perl-to-Python
transitions involve conceptual interference; the LDA result confirms
these sub-profiles are empirically distinguishable at the population
level.

\textbf{Topic 22 -- Scientific/Hardware Developer ($\sim$29k users).}
This profile spans Python (high-level orchestration), Fortran
(legacy numerical computing), and VHDL/Verilog (hardware
description). It captures developers who bridge software and
hardware, common in signal processing, scientific simulation, and
FPGA design. Farshidi~\cite{farshidi2021} showed language selection
depends heavily on domain context; this profile represents
developers for whom domain requirements force a combination unusual
outside that domain.

\textbf{Topic 14 -- Functional/Academic Programmer ($\sim$6k users).}
The smallest shown topic loads on Haskell, Lisp, Scheme, Rust, and
OCaml, languages sharing strong static typing, algebraic data
types, and formal reasoning support. Rust's presence alongside
Haskell and Lisp reflects its attraction to functional programmers
seeking systems-level type safety.

\textbf{Topic 23 -- Windows Enterprise Automation ($\sim$7k users).}
PowerShell and VBScript with near-zero weight on Bash or Shell
confirms a clean boundary between Windows-native and Unix-native
automation at the population level.

\textbf{Topic 11 -- Generic Web Developer ($\sim$66k users).}
The largest topic spreads weight broadly across PHP, JavaScript, and
Java without concentrating on any subset, the generic web developer
profile. Its diffuse distribution contrasts with all concentrated
topics above, indicating web developers use many overlapping
languages without a fixed dominant combination, consistent with
survey findings~\cite{durdev2024}.

\begin{figure}[h]
  \centering
  \includegraphics[width=\columnwidth]{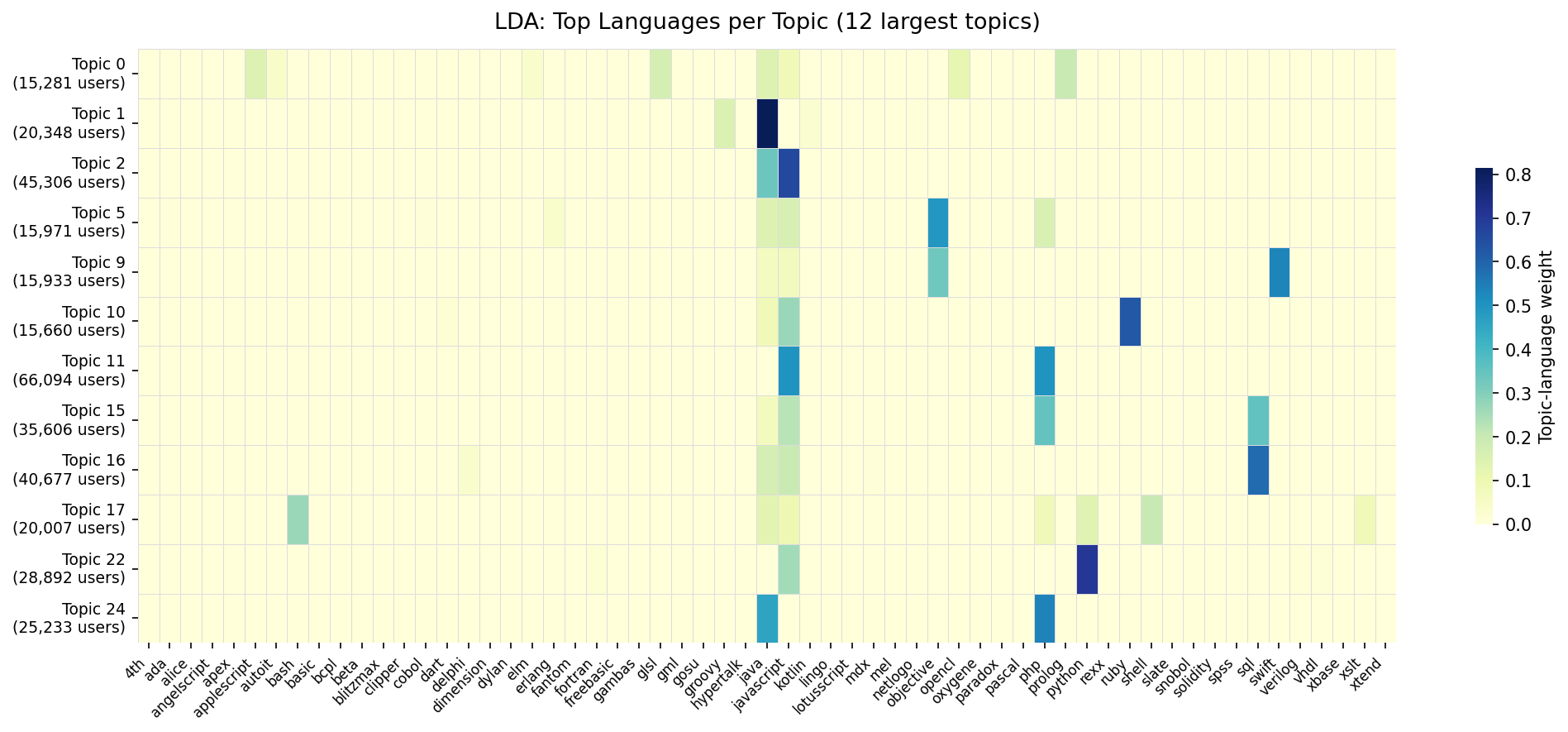}
  \caption{LDA topic-language weight heatmap (12 largest topics,
  top 8 languages each). Each row sums to approximately 1.
  Concentrated rows are domain-specific profiles; the diffuse
  row (Topic~11) is the generic web developer profile.}
  \label{fig:lda}
\end{figure}

Fig.~\ref{fig:lda} shows the topic-language weight heatmap for the
12 largest topics. Most topics show 1--3 bright cells and near-zero
elsewhere. Topic~11's broad distribution and Topic~9's near-total
concentration on swift/objective-c illustrate the extremes of the
profile spectrum.

\subsection{RQ3: Ecosystem Community Structure}
\label{subsec:rq3}

\textbf{Centrality.}
Table~\ref{tab:centrality} ranks languages by total co-usage weight,
capturing both breadth and intensity of co-occurrence.

\begin{table}[h]
  \caption{Top 10 languages by total co-usage weight}
  \label{tab:centrality}
  \begin{tabular}{lrr}
    \toprule
    Language & Degree & Total weight \\
    \midrule
    javascript  & 177 & 550,975 \\
    java        & 181 & 418,450 \\
    php         & 175 & 403,473 \\
    c\#         & 174 & 322,120 \\
    sql         & 173 & 299,412 \\
    python      & 177 & 287,996 \\
    c++         & 177 & 258,387 \\
    c           & 174 & 211,684 \\
    bash        & 163 & 115,720 \\
    ruby        & 158 & 110,066 \\
    \bottomrule
  \end{tabular}
\end{table}

Java has the highest degree (181 of 185 possible neighbors),
confirming the hub role identified in RQ1. JavaScript has the
highest total weight despite a lower degree, meaning its pairings
are more \emph{intense} rather than merely broader. Both occupy
central positions for different structural reasons: Java's
centrality is breadth-driven; JavaScript's is intensity-driven.
Bash's appearance in the top 10 reflects its role as a universal
infrastructure tool appearing across virtually all developer
profiles.

\textbf{Community Detection.}
Louvain yields modularity $Q = 0.096$ and three communities
(Table~\ref{tab:communities}).

\begin{table}[h]
  \caption{Louvain communities of the co-usage graph}
  \label{tab:communities}
  \begin{tabular}{clr}
    \toprule
    ID & Representative languages & Size \\
    \midrule
    0 & javascript, java, php, c\#, sql, ruby & 43 \\
    1 & objective-c, swift, applescript       & 10 \\
    2 & python, c++, c, bash, shell, perl, r  & 133 \\
    \bottomrule
  \end{tabular}
\end{table}

\textbf{Community 0 -- Web/Enterprise (43 languages).}
This community groups JavaScript, Java, PHP, C\#, SQL, Ruby, and
their smaller companions. It is the most populous and contains the
four highest-weight languages, but the modest lift values from RQ1
(1.0--1.5) confirm it is broadly coupled, not tightly specialized:
developers here tend to use many of these languages interchangeably
rather than pairing any specific two or three. This matches Moutidis
and Williams'~\cite{moutidis2021} finding that the web development
community is the largest and most persistent on Stack Overflow, with
users rarely moving to other communities.

\textbf{Community 1 -- Apple Ecosystem (10 languages).}
The smallest and most isolated community: all ten members are
Apple-platform languages. No cross-community edges exceed the
500-user threshold. This confirms that Apple-platform development
is a self-contained ecosystem with minimal overlap with the wider
developer population. The consistency across all three methods --
FP-Growth's highest lift, LDA's most concentrated topic, Louvain's
most isolated community, constitutes strong evidence of genuine
structural isolation.

\textbf{Community 2 -- Systems/Scientific (133 languages).}
The largest and most heterogeneous community groups C, C++, Python,
Bash, R, Fortran, VHDL, Verilog, and 123 additional languages,
unified by Unix/open-source heritage and performance or numerical
computing contexts. The graph detector merges them because
cross-usage density through Python and Bash is higher than the
Community~0/2 boundary density. However, Community~2 is not
homogeneous: LDA reveals at least two sub-profiles within it --
scientific/hardware computing (Topic~22) and Unix scripting
(Topics~3 and~17), connected by shared Python and Bash usage but
otherwise distinct.

\textbf{Interpreting the modularity value.}
The modularity $Q = 0.096$ is lower than in sparse social networks
(where $Q > 0.3$ is common) because most of the 186 languages
co-occur at some level across 435,803 users, limiting maximum
achievable $Q$. The three communities found represent the strongest
available topological signal, not an artifact.

\begin{figure}[h]
  \centering
  \includegraphics[width=\columnwidth]{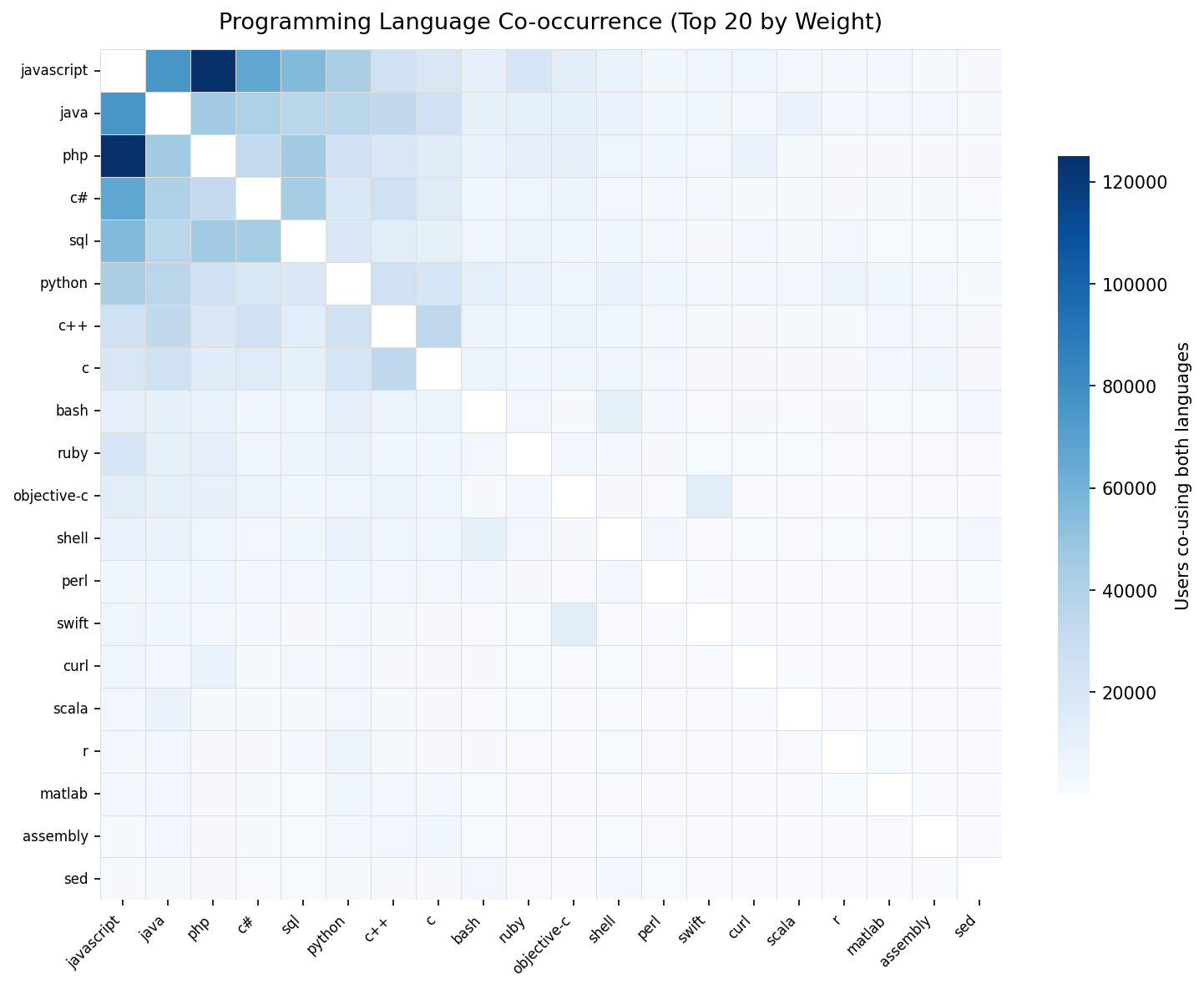}
  \caption{Co-occurrence heatmap (top 20 languages by weight).
  Darker cells indicate more shared users. Web/enterprise (upper
  left) and systems/scientific (lower right) are separated; C sits
  at the boundary.}
  \label{fig:heatmap}
\end{figure}

Fig.~\ref{fig:heatmap} visualizes the community structure. The
javascript/java/php block (Community~0, upper left) and the
python/c++ cluster (Community~2, lower right) are visually
separated, with C at the boundary. The Apple ecosystem languages
form a distinct cluster in the middle-left, well separated from
both main blocks.

\textbf{Synthesis: convergence across three methods.}
All three methods independently identify the same macro-structure.
FP-Growth's highest lifts correspond to the tightest community
boundaries. LDA's most concentrated topics correspond to the same
communities. Louvain assignments are consistent with both analyses.
This convergence across methods with different theoretical
assumptions: rule-based mining, probabilistic generative modeling,
and graph-theoretic partitioning, provides strong evidence that
the identified structure is real and stable.

% ------------------------------------------------------------------
\section{Discussion}

\subsection{What the Results Reveal About the Developer Ecosystem}

The most striking finding is the coexistence of radically different
types of coupling within a single ecosystem.
Shell/bash co-usage (lift 11.06) is \emph{functionally driven}: the
two tags refer to overlapping technology and are nearly definitional
synonyms in practice.
Swift/Objective-C coupling (lift 8.35) is \emph{platform-driven}:
Apple's developer environment enforces this pairing on any production
iOS/macOS codebase regardless of developer preference, and the
coupling will persist as long as any legacy Objective-C code remains
in production.
C/C++ coupling is \emph{ecosystem-driven}: the two languages
interoperate by design, share a runtime model, and are routinely
combined in systems programming to balance expressiveness and
performance.
Python and JavaScript, by contrast, exhibit near-independence despite
being the most prevalent languages, their coupling with any given
partner is roughly proportional to that partner's base rate.

This taxonomy, functional, platform, ecosystem coupling, and broad
adoption without coupling, provides a more nuanced characterization
of the language ecosystem than popularity rankings alone can offer.
Popularity measures which language is \emph{most used}; lift measures
which languages are \emph{used together beyond chance}. The two
capture fundamentally different structural properties, and conflating
them leads to incorrect conclusions about which language combinations
a tool or documentation set must support jointly.
Cherny-Shahar and Feitelson~\cite{cherny2024} found C to be a central
interoperability pillar at the project level; our user-level results
confirm and extend this by showing that C's role manifests both in
direct coupling with C++ and in cross-community hub status.

\subsection{Implications for Tool Builders and Language Designers}

The co-usage structure defines natural support boundaries for IDE and
tool developers that are grounded in observed developer behavior
rather than assumed use cases.
An IDE targeting Apple developers must handle Swift and Objective-C
together, treating either as an independently optional feature
would fail the majority of its intended users.
An IDE targeting scientific computing developers must anticipate
workflows spanning Python, Fortran, and hardware description
languages; siloing these languages into separate toolchain contexts
ignores how this population actually works.
Neither pairing is incidental, and the lift values quantify how
strongly the pairings hold across the developer population.

Java's high degree (181 of 185 possible neighbors) implies that
tools and documentation targeting Java must serve simultaneously
diverse audiences, web developers, systems programmers, enterprise
architects, whose secondary language contexts are entirely
different.
A Java IDE feature that makes sense for a web developer
(lightweight HTTP client integration) may be irrelevant to an
embedded developer using Java for firmware, and vice versa.
The LDA profiles provide an empirical segmentation that could inform
more targeted feature prioritization: rather than designing for
the average Java user, tool builders can design for the Apple,
systems, or web sub-population within their user base.

\subsection{Comparison to Prior Community Detection on Stack Overflow}

Moutidis and Williams~\cite{moutidis2021} applied community
detection to annual Stack Overflow user-technology interaction graphs
from 2008 to 2020, finding that web development is the dominant and
most persistent community, with limited cross-community user movement
over time.
Our Community~0 corroborates this at the language level and extends
it: the web/enterprise community is not only the most persistent but
also the most loosely coupled, with its members using a broad and
interchangeable set of languages rather than a fixed combination.
The key methodological difference is granularity and unit of
analysis: their framework-inclusive, temporally-sliced approach
operates on all technology tags and tracks community membership
year by year, whereas ours focuses exclusively on programming
languages and operates on sustained per-user co-usage.
This difference explains why the clean Apple ecosystem isolation
(Community~1) was not a prominent finding in their analysis --
including framework and tool tags such as Xcode, CocoaPods, and
UIKit alongside the language tags dilutes the Apple-specific
language signal across a much larger tag space, making the
community less distinct.
Our language-only focus sharpens this boundary considerably, and
the convergence of all three independent methods on the same
isolated community strengthens confidence that it reflects a
genuine structural reality rather than a methodological artifact.

\subsection{Stack Overflow in the Age of Generative AI}

The behavioral data underpinning this study was generated in an era
when Stack Overflow was the primary destination for developer
problem-solving. That context is shifting rapidly. Del Rio-Chanona
et al.~\cite{delriochanona2023} documented a 25\% decline in Stack
Overflow activity within six months of ChatGPT's release, with the
largest drops concentrated in posts about the most widely used
programming languages, precisely the languages at the center of
the co-usage graph studied here. Burtch et al.~\cite{burtch2024}
found that the decline is concentrated among newer users who lack
deep social embeddedness in the platform, while more experienced
contributors remain active, effectively raising the average
expertise level of remaining posts. Gallea~\cite{gallea2023}
similarly found that questions posted after ChatGPT's launch are
more complex and better documented, suggesting that routine
problem-solving is migrating to LLM assistants while the harder,
more context-specific questions remain on the forum.

These trends raise a legitimate question about the long-term
viability of Stack Overflow as a behavioral data source for
research. If activity continues to decline, particularly among
junior developers who generate incidental co-usage, the sustained
multi-language population that our filtering step targets may
become an increasingly expert and unrepresentative slice. Future
iterations of this pipeline applied to post-2023 data would capture
a community self-selected for complexity and depth, which could
shift the co-usage structure in ways that a static snapshot cannot
reveal.

At the same time, the forum's value as a \emph{research artifact}
is arguably not diminished by declining activity, it may even be
enhanced. Helic et al.~\cite{helic2025} showed that Stack Overflow
is not obsolete: the remaining questions are harder, longer, and
carry richer signal than before. For the purposes of mining
behavioral patterns, a smaller but more deliberate and expert
population may produce cleaner co-usage traces than the
historically larger but noisier one. The cumulative archive --
spanning over fifteen years of developer activity, remains the
largest available record of multi-language developer behavior
and cannot be replicated by any other platform or LLM interaction
log. For studying \emph{how developers combine technologies}, as
opposed to \emph{how they resolve individual errors}, the forum's
historical depth continues to make it uniquely valuable.

\subsection{The Convergence Result}

The methodologically most significant finding is that FP-Growth
(rule-based, no probabilistic assumptions), LDA (probabilistic
generative model, no graph structure), and Louvain (graph-theoretic,
no language-level model) converge on the same macro-structure.
The Apple ecosystem's isolation, the shell/bash tight coupling,
the C/C++ pairing, Java's hub status, and the web/enterprise
versus systems/scientific split all appear consistently across
all three methods.
This convergence matters precisely because each method has different
theoretical assumptions and different failure modes: LDA is sensitive
to the choice of $k$; FP-Growth is sensitive to support and
confidence thresholds; Louvain is sensitive to graph density and
can be non-deterministic.
The fact that all three agree on the same structure substantially
reduces the probability that any finding is an artifact of any
single method's assumptions or parameter choices, providing a
stronger form of empirical validation than any one method could
supply alone.
In practice, this means a tool builder or language designer can
act on these findings with greater confidence: the ecosystem
structure they describe is not a statistical curiosity that
disappears under different modeling choices.

% ------------------------------------------------------------------
\subsection{Threats to Validity}

\textbf{Population representativeness.}
Stack Overflow over-represents web, mobile, and systems developers
and under-represents those working in proprietary or
non-English-speaking environments.
Languages common in aerospace, defense, embedded systems, or
scientific computing domains that are served by specialized
forums may appear peripheral or absent from the co-usage graph,
even if they are heavily used in practice.

\textbf{Tag noise and usage signal validity.}
The analysis treats tagged activity as evidence of language
familiarity, an assumption that is violated in two directions:
users who ask questions about languages they do not know, and
users who answer questions beyond their primary expertise.
The curated language list removes 76 non-language tags but cannot
eliminate all noise, and languages with ambiguous tag names may
be systematically over- or under-represented.
Future work could partially address this by requiring a minimum
post score or accepted-answer threshold as an additional signal
of genuine expertise.

\textbf{Filtering effects.}
The 82\% filter emphasizes sustained multi-language practitioners
and excludes novice or casual users whose co-usage patterns may
differ substantially.
If novice developers systematically start with one language and
only gradually adopt a second, the pipeline by construction
excludes their early-career phase, biasing all three analyses
toward more experienced practitioners.
The results should therefore be interpreted as characterizing the
sustained multi-language developer population rather than the full
range of Stack Overflow users.

\textbf{Static analysis.}
Aggregating all posts through 2024 into a single snapshot ignores
temporal dynamics.
The rise of TypeScript, the sustained decline of Objective-C since
Swift's 2014 introduction, and the emergence of Rust as a systems
language are invisible in the aggregate, and co-usage relationships
that were strong in 2010 but have since weakened may still appear
in the results.
Year-segmented analysis is a natural and important extension that
would reveal whether the three-community macro-structure is stable
over time or reflects a transitional state.

\textbf{Methodological sensitivity.}
LDA results depend on the number of topics $k$; FP-Growth depends
on the support $\sigma$ and confidence $\gamma$ thresholds; and
Louvain is non-deterministic, though it produced the same
three-community macro-structure across all five independent runs.
More permissive FP-Growth thresholds would surface weaker rules
and could obscure the signal-to-noise ratio, while a stricter
minimum support would eliminate lower-frequency but potentially
meaningful pairings.
These sensitivities do not invalidate the findings but warrant
caution in applying the derived profiles to rapidly evolving
or highly specialized ecosystems where the current thresholds
may not be well-calibrated.
% ------------------------------------------------------------------
\section{Conclusion and Future Work}

We presented a three-method empirical analysis of programming
language co-usage patterns on Stack Overflow, constructed from
36 million posts across 435,803 sustained multi-language developers
and 186 languages. The three RQs are answered as follows.
\textbf{RQ1:} Co-usage coupling varies from near-definitional
(shell/bash, lift 11.06) through platform-enforced (Swift/
Objective-C, lift 8.35) to ecosystem-driven (C/C++, lift 3.45)
and near-independence (Python, JavaScript). Lift effectively
separates structural coupling from popularity effects.
\textbf{RQ2:} Twenty-five developer profiles emerge, including
Apple-platform developers, scientific/hardware computing,
functional/academic programming, Windows enterprise automation,
and two distinct Unix scripting sub-profiles. The generic web
developer is the largest ($\sim$66k users) but most diffuse
profile. \textbf{RQ3:} The ecosystem partitions into three
macro-communities, web/enterprise (43 languages), Apple ecosystem
(10 languages), and systems/scientific (133 languages), with
Java as the highest-degree hub bridging all three.

The convergence of all three methods on the same macro-structure
is the most methodologically significant result, advancing prior
MSR work by Moutidis and Williams~\cite{moutidis2021} and
Cherny-Shahar and Feitelson~\cite{cherny2024} by operating at the
per-user co-usage level with three complementary analyses.

The findings presented here are descriptive and static, and several
directions emerge naturally from their limitations. The co-usage
structure, developer profiles, and community boundaries are all
derived from a single aggregate snapshot of developer behavior,
 one that predates the significant shifts in platform activity
documented in the preceding discussion. As Stack Overflow's
population evolves under the influence of generative AI tools,
the ecosystem structure it encodes will evolve with it, and the
profiles derived today may not reflect the community of tomorrow.
Extending and validating this work therefore requires both temporal
depth and cross-platform breadth, alongside downstream studies that
move from describing the ecosystem to acting on the description.

\textbf{Temporal analysis.}
Year-segmented slices (e.g., two-year windows from 2010 to 2024)
would reveal how quickly the Swift/Objective-C coupling weakens
as legacy code is migrated, whether Rust is drawing developers from
the C/C++ cluster, and whether TypeScript is consolidating or
fragmenting JavaScript usage patterns.

\textbf{Enriched user signals.}
Incorporating user reputation, post score, and acceptance rate as
weighting signals would distinguish incidental from expert-level
co-usage. Adding framework tags (React, Django, Spring) would
enable finer-grained profiling within the web/enterprise community.

\textbf{Cross-platform validation.}
GitHub activity and developer survey data provide complementary
signals to confirm whether co-usage patterns on Stack Overflow
reflect actual codebase composition or are artifacts of the
question-answering context.

\textbf{Application to programming education.}
The two-profile Unix scripting finding has direct curriculum
implications: designers must decide whether Perl-heavy and
Python-heavy scripters are a unified audience or two distinct
learner populations with different background knowledge. Wu et
al.~\cite{wu2023} linked Stack Overflow knowledge to programming
textbook chapters; our profiles could identify which developer
archetypes are underserved by existing educational materials.

\textbf{Downstream application studies.}
The co-usage structure could serve as input to tool adoption
recommendation, documentation quality assessment, and developer
team composition studies. Validating the profiles in these applied
contexts would transform the current descriptive contribution into
a practically evaluated analytical resource.

% ------------------------------------------------------------------
\bibliographystyle{ACM-Reference-Format}
\bibliography{refs_msr}

\end{document}